\documentclass[sigconf]{acmart}
\usepackage{tabularx}
\usepackage{subcaption}
\usepackage{multirow}
\usepackage{booktabs}
\usepackage{amsmath}
\usepackage{subcaption}
\usepackage{longtable}
\usepackage{array}

\copyrightyear{2025}
\acmYear{2025}
\setcopyright{cc}
\setcctype{by}
\acmConference[FDG '25]{International Conference on the Foundations of Digital Games}{April 15--18, 2025}{Graz, Austria}
\acmBooktitle{International Conference on the Foundations of Digital Games (FDG '25), April 15--18, 2025, Graz, Austria}
\acmPrice{}
\acmDOI{10.1145/3723498.3723846}
\acmISBN{/25/04}

\begin{document}
\title{Analysis of Uncertainty in Procedural Maps in Slay the Spire}

\author{Mahsa Bazzaz}
\affiliation{%
  \institution{Northeastern University}
  \city{Boston, Massachusetts}
  \country{USA}}
\email{bazzaz.ma@northeastern.edu}
\orcid{0009-0004-0022-9611}

\author{Seth Cooper}
\affiliation{%
  \institution{Northeastern University}
  \city{Boston, Massachusetts}
  \country{USA}}
\email{se.cooper@northeastern.edu}
\orcid{0000-0003-4504-0877}

\begin{abstract}
This work investigates the role of uncertainty in Slay the Spire using an information-theoretic framework. Focusing on the entropy of game paths (which are based on procedurally-generated maps) we analyze how randomness influences player decision-making and success. By examining a dataset of 20,000 game runs, we quantify the entropy of paths taken by players and relate it with their outcomes and skill levels. The results show that victorious runs are associated with higher normalized entropy, suggesting more risk-taking. Additionally, higher-skill players tend to exhibit distinct patterns of risk-taking behavior in later game stages. 
\end{abstract}

\maketitle

\section{Introduction}
Slay the Spire\citep{slaythespiregame} is a strategic deck-building game where players navigate a procedurally generated\citep{togelius2011procedural} path, engaging in battles using a customizable deck of cards. Players select one of several unique characters, each with their own abilities and card sets. Success in the game depends on how well players manage their resources, plan their moves, and adapt to evolving challenges. The ultimate goal is to defeat powerful bosses and, eventually, reach the final boss.

Each attempt at the game, known as a ``run'', begins when the player starts the game and ends either in victory or failure. If the player fails, they must begin a new run from the start. This is because this game is a roguelike game, meaning that the game is designed with the expectation that players will fail and restart multiple times. This iterative process allows players to enhance their skills, deepen their understanding of the game, and unlock new content, such as additional cards, characters, and relics. These improvements and unlocked elements increase the likelihood of success in subsequent runs.

Randomness plays a significant role in digital games, ranging from simple mechanics like rolling dice to more complex behaviors in AI systems. These elements, though often outside the player's control, have a substantial impact on gameplay experience and outcomes. Previous studies have explored the effects of randomness in Slay the Spire, particularly in relation to cards and relics, showing that higher-skilled players manage these random elements better than lower-skilled players. However, players across all skill levels are still influenced by the inherent randomness of the cards and relics~\cite{andersson2023you,lantz2017depth}.

\begin{figure}[h]
\includegraphics[width=0.3\textwidth]{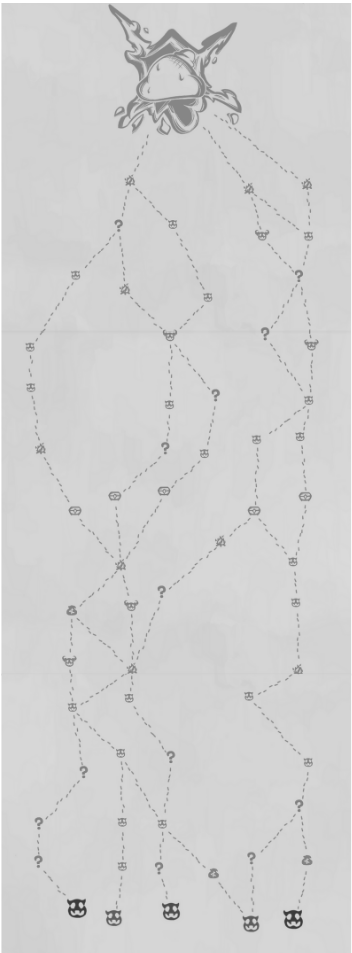}
\caption{Screenshot of the game showcasing the path selection in Slay the Spire.}
\label{fig:path}
\end{figure}

Despite this, the randomness of an entire ``run'' in Slay the Spire has not been thoroughly examined. This is particularly important because players are presented with multiple paths, each offering different map locations and varying levels of uncertainty. This aspect is especially significant in Slay the Spire, as its procedurally generated maps introduce numerous possible layouts and choices --- far more than would be found in a game with manually designed maps. Choosing a specific path allows players to adjust the game's difficulty by encountering easier rooms or taking advantage of resources like money, campfires for healing, or strengthening to face elite enemies. Figure \ref{fig:path} shows a sample path selection in Slay the Spire. The paths are shown vertically in the original game.

Information theory has been previously applied to game research to understand difficulty \cite{chen2023entropy} and depth \cite{lantz2017depth} of games, and the evolution of the game’s system over time\cite{zuparic2021information}.

However, no studies to date have analyzed the gameplay of Slay the Spire using information theory.

This work aims to fill that gap by examining how randomness and entropy influence player behavior and outcomes in Slay the Spire. By applying information theory, we seek to understand how players confront and manage randomness throughout a run and how these elements related with their success.

Our research questions are:

\textbf{RQ1}: Which runs (ended with Victory or Defeat) involve more uncertain paths?

\textbf{RQ2}: Which type of player (lower-level or higher-level) tends to choose more uncertain paths?

Our contributions are summarized as follows:
\begin{itemize}
\item We define the concepts of ``Uncertainty'' and ``Risk-Taking'' within the context of the game Slay the Spire, and propose an information-theoretic approach to quantify these concepts.
\item We analyze the uncertainty associated with all potential paths available to players using a comprehensive set of player logs.
\item We investigate the relationship between path uncertainty and player success in game runs, assessing whether risk-taking strategies provide an advantage in Slay the Spire.
\item We explore the relationship between path uncertainty and player ascension level in runs, identifying which types of players are more inclined towards risk-taking strategies.
\end{itemize}


\section{Related Work}
\subsection{Information Theory}

Information theory has become an instrumental tool in understanding the dynamics, complexity, and engagement levels within games. In strategic and puzzle-based games, it provides a framework to quantify uncertainty, player choices, and the structural characteristics that define a game’s complexity and difficulty.

\citet{lantz2017depth} examine the concept of ``depth'' in strategic games through information theory, aiming to define depth as a measurable property reflecting a game’s capacity for prolonged engagement and continuous learning. They propose a metric, depth, corresponding to a game's susceptibility to computational exploration, effectively capturing the level of challenge and variation a game presents as players encounter new strategies. This depth is often associated with games that foster enduring interest, as they offer continuous layers of strategy that emerge over time.

\citet{chen2023entropy} apply entropy to single-player puzzle games, focusing on how uncertainty in solving puzzles relates to difficulty. They introduce entropy-based algorithms to measure a puzzle's uncertainty, which corresponds to the number of choices or steps required for solution discovery. By calculating entropy scores for each puzzle state, they assess the difficulty level based on the choices available to players at each step, ultimately creating a metric that predicts player challenge and engagement with different puzzles.

\citet{zuparic2021information} leverage Shannon entropy and information storage concepts in the analysis of archetypes in the collectible card game Hearthstone. By measuring the entropy of player choices over time, they assess the game’s adaptability and complexity as players experiment with new strategies. Their study shows that higher entropy values correlate with phases of player exploration, particularly during periods of significant game updates or shifts in available strategies. This dynamic analysis provides insights into the game’s ``meta'' and its evolution in response to player behavior and system changes.

More recently, \citet{shen2024generalized} expanded on their prior work \citep{chen2023entropy}, by introducing generalized entropy and solution information to assess puzzle difficulty. They propose novel metrics, Minimum Solution Information (MSI) and Total Solution Information (TSI), to quantify the information required for players to solve puzzles, considering their knowledge and policies. These methods enhance earlier work by providing a mathematically grounded approach for evaluating player engagement and difficulty in puzzle games such as The Witness. Their findings highlight how entropy-based measures can predict player satisfaction and guide the design of adaptive puzzle curricula.

\subsection{Slay the Spire}

Slay the Spire is a popular roguelike deck-building game known for its strategic complexity, where randomness plays a critical role in shaping the player experience. The game’s mechanics, including card draws, enemy encounters, and item rewards, create an unpredictable environment, requiring players to adapt dynamically to changing circumstances. Studies on Slay the Spire emphasize that while randomness can enhance the game’s replayability and challenge, it also introduces elements of luck that impact player success.

\citet{nam2023exploring} discuss how randomness in Slay the Spire such as the order of card draws and the choice of routes on the map, significantly influences gameplay dynamics. Their analysis shows that randomness can lead to both satisfaction and frustration, as players must navigate unexpected scenarios and adapt their strategies accordingly. They propose frameworks for developing pseudorandom systems that balance unpredictability with fairness, ensuring an engaging but not overly punishing experience.

\citet{andersson2023you} explore how elements of randomness, or EoR, differentiate player skill levels. Their findings reveal that high-skilled players tend to embrace randomness more, utilizing cards with unpredictable effects to gain strategic advantages. This study suggests that skilled players view randomness as an opportunity for creative problem-solving, rather than a hindrance, indicating that familiarity with the game's systems allows for better adaptation to unexpected outcomes. They highlight the complexity of balancing randomness to maintain player engagement without reducing perceived fairness.

These studies highlight the nuanced role of randomness in Slay the Spire. However, no studies to date have analyzed the ``uncertainty'' of Slay the Spire paths and the concept of ``Risk-Taking'' using information theory.

\section{Terminology}
This section will describe terminology specific to the domain of Slay the Spire that will be used throughout this paper. All definitions have been exactly quoted from \citet{andersson2023you}.

\textbf{Act} - ``An act describes a section of the gameplay. It is the way Slay The Spire describes chapters in the game. The game consists of three default acts and one bonus act.''

\textbf{Ascension} - ``This is the hard mode of the game. A higher ascension level means a more difficult run as each level adds a negative modifier to your gameplay.''

\textbf{Boss} - ``Bosses are incredibly difficult monsters that are encountered at the end of each Act.''

\textbf{Elite} - ``Elites are stronger monster encounters that are marked separately from normal ones.''

\textbf{Floor X} - ``A quantitative description for encounters. A floor holds a numeric value X and represents how far the player has reached the game.''

\textbf{Monster} - ``Monsters are the basic enemies of Slay the Spire.''

\textbf{Potion} - ``Potions are one-time-use items that can only be used in combat.''

\textbf{Relic} - ``These are items that have a permanent effect on your entire instance of the game (run). Each character starts an instance of a game with a specific relic.''

\textbf{Run} - ``The gameplay loop spanning from a new start until the last boss has been defeated or the player has been defeated.''

We also define ``Uncertainty'' and ``Risk-Taking'' within the context of Slay the Spire:

\textbf{Uncertainty} - Some map locations in Slay the Spire exhibit more uncertainty than others, meaning they are more unpredictable compared to other locations. This means that before entering these locations, players face a higher degree of uncertainty about the challenges or events they will encounter.

\textbf{Risk-Taking} - We define Risk-taking as the willingness of the player to choose paths with more ``Uncertainty''. In Slay the Spire, harder fights often lead to better rewards. So, players who take more risks are willing to face tougher challenges to earn bigger rewards.

\section{Dataset}
In this study, we utilize log data released by the developers of the game, which contains 77 million detailed game runs \cite{slay_the_spire_2020}. The attributes used in our analysis are listed in Table \ref{tab:variable}. To ensure comparability between game runs, we focus on logs with fixed attributes listed in Table \ref{tab:fixed}, which account for consistent game settings across runs.

Two key attributes, the Boolean variable \texttt{Victory} and the integer variable \texttt{Ascension Level}, are used to categorize players based on their experience level and success in the run.

Ascension is a game mode in Slay the Spire designed to increase difficulty by adding various modifiers. This mode consists of 20 cumulative levels, where each level inherits the modifiers of the previous levels. Higher Ascension levels correspond to increased difficulty, making them an effective measure of a player’s skill and experience.

The final two attributes, \texttt{Seed played} and \texttt{Path taken}, are crucial for our analysis. The \texttt{Seed played} variable helps us to regenerate complete maps, while \texttt{Path taken} captures the choices a player made in selecting their path through the game.

We utilize \textbf{the Official Wiki Fandom of Slay the Spire}\footnote{\url{https://slay-the-spire.fandom.com/wiki/Slay_the_Spire_Wiki}} as a key resource in this work. This comprehensive, community-driven Wiki, hosted on the Fandom platform, serves as an authoritative repository for Slay the Spire, providing detailed information on gameplay mechanics, characters, cards, relics, and strategies. The content is created and maintained by fans, with active collaboration from the game developers, who supply official game assets and information to ensure its accuracy and relevance.

In particular, we gathered all data regarding the probability of each map location from this resource.

\section{Methodology} \label{methodology}
We would like to explore the uncertainty of paths by measuring the entropy\citep{renyi1961measures} of each map location (Equation \ref{eq:entropy1}) based on the probability distribution of each map location type. There are 6 types of map locations in Slay the Spire, detailed in Table \ref{tab:map_sites}.

\begin{equation}
\text{Location Entropy} = -\sum p_i \log p_i
\label{eq:entropy1}
\end{equation}

We calculate the entropy contribution of each map location based on its type, act, and frequency of visits. By accumulating the entropies of these locations, we quantify the overall path entropy.

After that, we find all possible paths the player could have taken based on the map, and we calculate the entropies of all these possible paths the player could have taken, and we determine the minimum and maximum values, which represent the range of entropy that was available to the player.

This step is crucial to make calculated entropies between different runs (and therefore with different minimum and maximum possible entropies) comparable. As shown in Equation \ref{eq:normalized_entropy}, we normalize the entropy of the chosen path, scaling it to a value between 0 and 1 according to its respective minimum and maximum range.

\begin{multline}
\text{Normalized Entropy} = \\
\frac{\text{Played Path Entropy} - \text{Minimum Possible Path Entropy}}{\text{Maximum Possible Path Entropy} - \text{Minimum Possible Path Entropy}}
\label{eq:normalized_entropy}
\end{multline}

Defeated runs can be incomplete due to the player's death, so all possible paths in these runs are calculated between the starting point of the map and the location where the player have died. As these paths can be shorter than complete paths in victorious runs, comparing entropy between defeated and victorious runs requires special consideration. To address this, as shown in Equation \ref{eq:average_normalized_entropy} we compare the \textbf{average entropy per step} across all runs, by dividing normalized entropy by the path length.

\begin{equation}
\text{Average Normalized Entropy Per Step} = \frac{\text{Normalized Entropy}}{\text{Path Length}}
\label{eq:average_normalized_entropy}
\end{equation}

\begin{table*}[h!]
\centering
\begin{tabular}{|p{0.3\linewidth}|p{0.6\linewidth}|}
\hline
\textbf{Location} & \textbf{Description} \\
\hline
\textbf{Unknown Location} & Unknown Room. Often contains an event, but may also contain a non-Elite enemy encounter, Treasure Room, or Shop. The likelihoods of these options are changed by \textit{Juzu Bracelet} and \textit{Tiny Chest}. \\
\hline
\textbf{Shop} & Buy cards, relics, potions, and/or remove a card from your deck, at the cost of some Gold. \\
\hline
\textbf{Treasure Room} & Contains a chest with 1 random Relic and potentially Gold. Affected by \textit{N'loth's Hungry Face}, \textit{Cursed Key}, and \textit{Matryoshka}. \\
\hline
\textbf{Rest Site} & Rest (heal HP) or Smith (upgrade a card). \textit{Shovel}, \textit{Girya}, and \textit{Peace Pipe} add additional options. \\
\hline
\textbf{Enemy} & Basic monster fights (non-Elites). \\
\hline
\textbf{Elite} & Fight an Elite monster, gaining a relic as a reward in addition to normal bonuses. (With the \textit{Black Star} you gain 2 relics.) An animated flame behind the Elite represents a buffed Elite that holds the Emerald Key for Act 4. \\
\hline
\textbf{Boss} & The end of each map leads to a boss. The reward is a choice of 1 of 3 Rare Cards and a choice of 1 of 3 Boss Relics. \\
\hline
\end{tabular}
\caption{Various map locations and their descriptions. Sourced from \url{https://slay-the-spire.fandom.com/wiki/Map_locations}.}
\label{tab:map_sites}
\end{table*}

\section{Experiments}

\begin{figure*}[t]
\centering
\begin{tabular}{cc}
\subcaptionbox{Normalized Entropies for Act 1}[0.5\textwidth]{\includegraphics[width=0.5\textwidth]{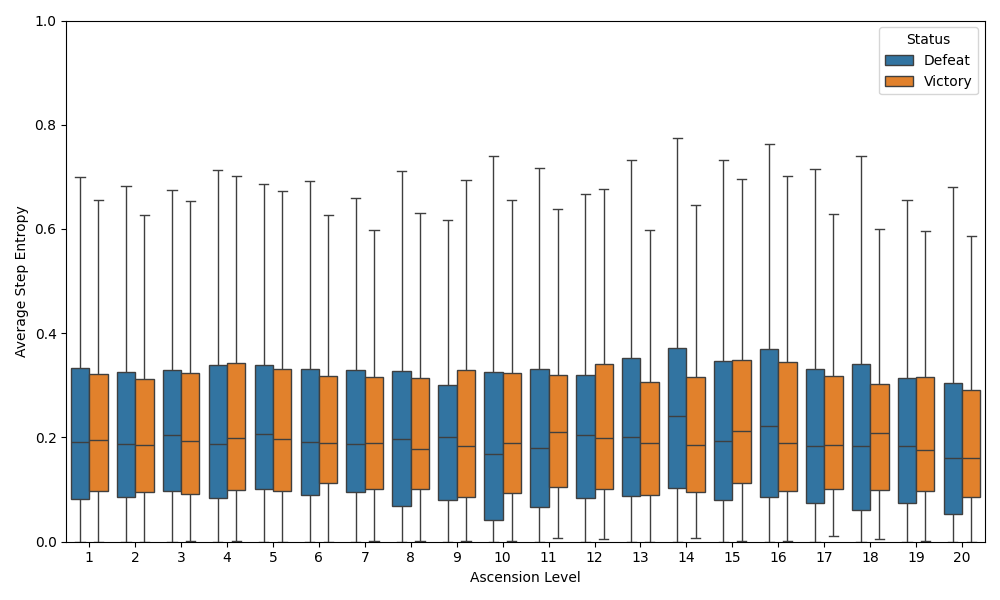}} &
\subcaptionbox{Normalized Entropies for Act 2}[0.5\textwidth]{\includegraphics[width=0.5\textwidth]{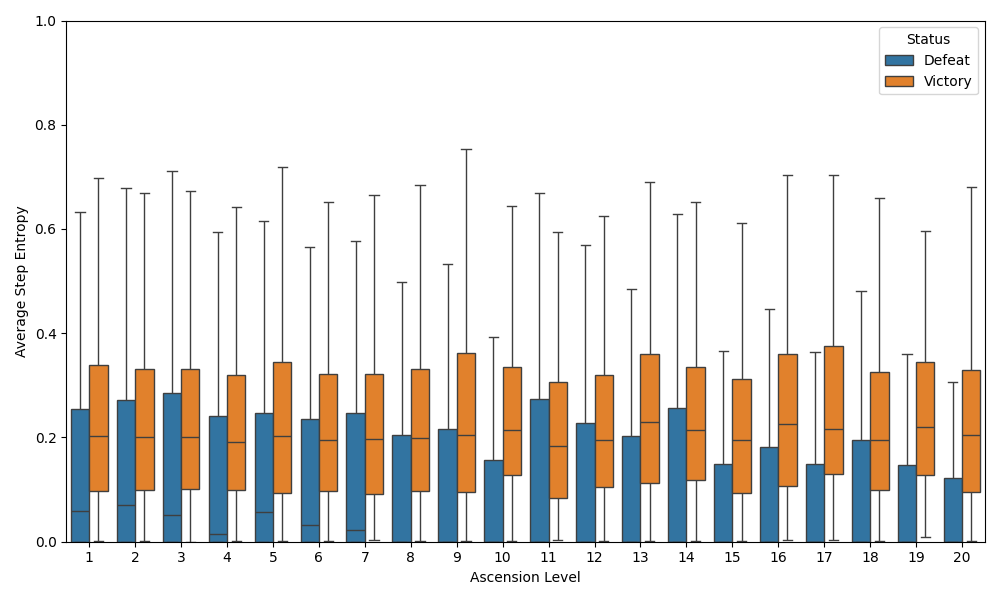}} \\
\subcaptionbox{Normalized Entropies for Act 3}[0.5\textwidth]{\includegraphics[width=0.5\textwidth]{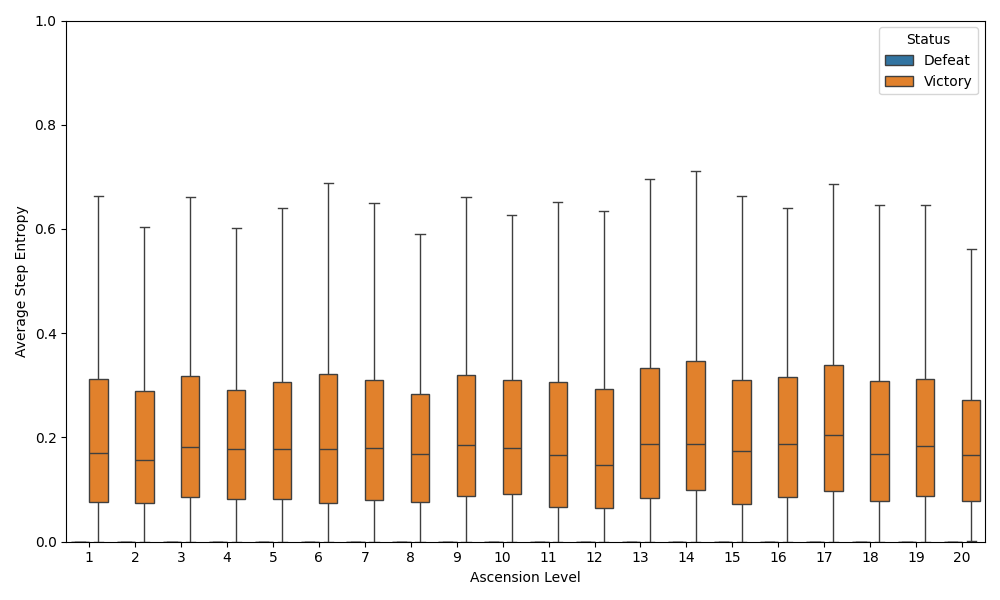}} &
\subcaptionbox{Average Normalized Entropies for all Acts}[0.5\textwidth]{\includegraphics[width=0.5\textwidth]{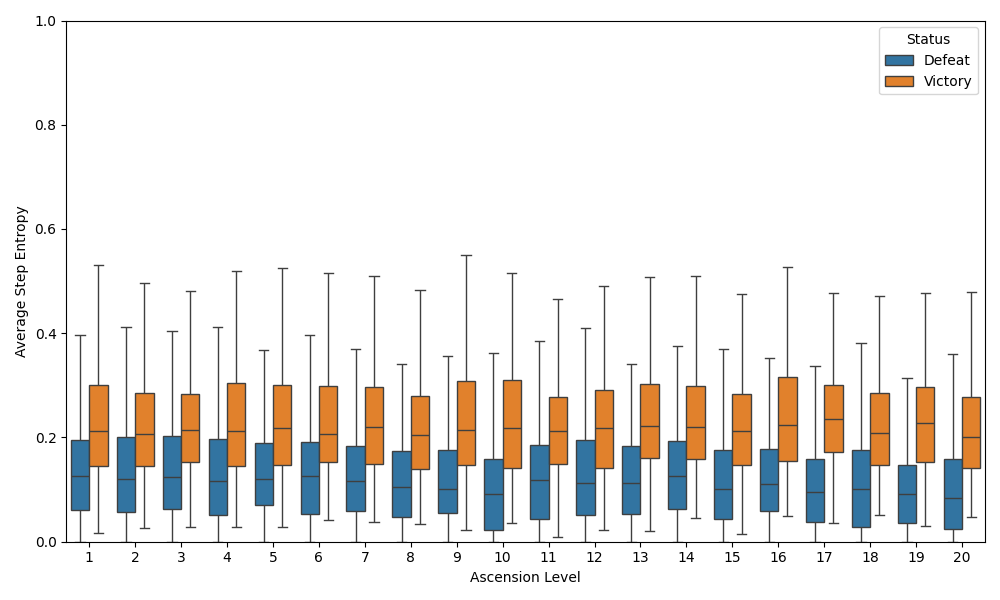}} \\
\end{tabular}
\caption{Relationship between Normalized Entropy and End Result of the Run}
\label{fig:results}
\end{figure*}
\subsection{Parsing the Data}
We begin by filtering the dataset of runs to ensure that we only accumulate samples with comparable game settings. Specifically, we select runs that are in ascension mode, excluding trial, daily, beta, and endless modes, and the runs with a specific chosen seed. We also only focus on those runs played with the Iron Clad character. This set of parameters ensures that the runs are played under the same game mode and with the same character. From this filtered dataset, we then randomly sample 20,000 runs for further analysis.

\subsection{Generating the Full Maps}
The logs in this dataset only include the path taken by the player and do not provide the complete map encountered during the run. Having data about the player's chosen path without knowing the other possible paths offers limited insights. To address this, we leveraged previous work that developed an oracle capable of perfectly replicating the map generation process of Slay the Spire\footnote{\url{https://github.com/Ru5ty0ne/sts_map_oracle}}.  Using the seed number from each run, we generate the complete map the player encountered with this oracle. To ensure the generated maps from this outsourced map generation replica are accurate, we verify that the player's path matches the generated map. If the paths do not align, we discard the run from the dataset.
 
\subsection{Generating All the Possible Paths}
We represent the map as a graph and we determine all unique paths in a graph between specific sets of starting and ending nodes by leveraging depth-first traversal. The graph is represented as an adjacency list, mapping each node to its list of connected neighbors. Paths are explored recursively, starting from a given node and extending to its unvisited neighbors until the target node is reached. A set tracks nodes in the current path, preventing cycles and ensuring each path is valid. If the starting node matches the target, the accumulated path is returned as a result. The code systematically explores and combines paths by iterating through all combinations of starting and ending nodes, removing duplicates to produce a final list of unique paths. This method provides a comprehensive approach to pathfinding, ensuring every possible route between the specified nodes is accounted for.

\subsection{Uncertainty of Monster (Enemy) Rooms}
Monsters serve as the ``basic enemies'' in Slay the Spire. Encountering their icon on the map triggers a battle against one or more, depending on the floor. Defeating them rewards you with coins, a choice of three cards, and occasionally a potion\footnote{\url{https://slay-the-spire.fandom.com/wiki/Monsters}}.

\subsubsection{Act 1: First Three Encounters}
Based on the probability of encounters in Table \ref{tab:act1_first} we have:

\begin{multline*}
    H(\text{Act 1 first 3 Monster encounters}) = -\sum p_i \log p_i \\
    = -\left[(25\% \log 25\%) + (25\% \log 25\%) + (25\% \log 25\%) + (25\% \log 25\%)\right] \\
    = -\left[4 \times (25\% \log 25\%)\right]
    = 4 \times 0.5 = 2
\end{multline*}

\subsubsection{Act 1: Rest of Encounters}
Based on the probability of encounters in Table \ref{tab:act1_rest} we have:

\begin{multline*}
H(\text{Act 1 rest of Monster encounters}) = -\sum p_i \log p_i \\
= -\Big[(6.25\% \log 6.25\%) + (12.5\% \log 12.5\%) + (6.25\% \log 6.25\%) + \nonumber \\
 (12.5\% \log 12.5\%) + (6.25\% \log 6.25\%) + (12.5\% \log 12.5\%) + \nonumber \\
 (12.5\% \log 12.5\%) + (9.375\% \log 9.375\%) + \nonumber \\
 (9.375\% \log 9.375\%) + (12.5\% \log 12.5\%)\Big] \nonumber \\
= -\Big[3 \times (6.25\% \log 6.25\%) + 5 \times (12.5\% \log 12.5\%) + \nonumber 2 \times (9.375\% \log 9.375\%)\Big] \nonumber = \\
3.265
\end{multline*}

\subsubsection{Act 2: First Two Encounters}
Based on the probability of encounters in Table \ref{tab:act2_first} we have:

\begin{multline*}
H(\text{Act 2 first 2 Monster encounters}) = -\sum p_i \log p_i \\
= -\left[ 5 \times (20\% \log 20\%) \right] \\
= -\left[ 5 \times (0.2 \times -2.32) \right] = -5 \times -0.46 = 2.32
\end{multline*}

\subsubsection{Act 2: Rest of Encounters}
Based on the probability of encounters in Table \ref{tab:act2_rest} we have:

\begin{multline*}
H(\text{Act 2 rest of Monster encounters}) = -\sum p_i \log p_i \\
= -\big[2 \times (7\% \log 7\%) + 3 \times (10\% \log 10\%) \\
\quad + (14\% \log 14\%) + 2 \times (21\% \log 21\%)\big] = 2.87
\end{multline*}

\subsubsection{Act 3: First Two Encounters}
Based on the probability of encounters in Table \ref{tab:act3_first} we have:

\begin{multline*}
H(\text{Act 3 first two monster encounters}) = -\sum p_i \log p_i \\
= -\big[3 \times (33.33\% \log 33.33\%)\big] = 1.58
\end{multline*}

\subsubsection{Act 3: Rest of Encounters}
Based on the probability of encounters in Table \ref{tab:act3_rest} we have:

\begin{multline*}
H(\text{Act 3 rest of monster encounters}) = -\sum p_i \log p_i \\
= -\big[8 \times (12.5\% \log 12.5\%)\big] = 3.0
\end{multline*}

\subsection{Uncertainty of Elite Rooms}
Elites are more powerful enemies distinguished from regular monsters on the map. Defeating an elite rewards you with a random relic, 25-35 gold, a card choice, and a score bonus. While their relics can be extremely valuable, engaging them may not always be wise, depending on the strength of one's deck. The same elite can appear multiple times within an act, and have an equal chance of appearing, but will never be encountered consecutively\footnote{\url{https://slay-the-spire.fandom.com/wiki/Elites}}.

\begin{itemize}
    \item \texttt{Gremlin Nob}, \texttt{Lagavulin} and \texttt{Three Sentries} are Act 1 Elites.
    \item \texttt{Book of Stabbing}, \texttt{Gremlin Leader} and \texttt{Taskmaster} are Act 2 Elites.
    \item \texttt{Giant Head}, \texttt{Nemesis} and \texttt{Reptomancer} are Act 3 Elites.
\end{itemize}

\subsubsection{First Encounter}
\begin{multline*}
H(\text{First elite encounter}) = -\sum p_i \log p_i \\
= -\big[3 \times (33.33\% \log 33.33\%)\big] = 1.58
\end{multline*}
\subsubsection{Rest of Encounters}
\begin{align*}
H(\text{Rest of elite encounters}) &= -\sum p_i \log p_i \\
&= -\big[2 \times (50\% \log 50\%) + (0\% \log 0\%)\big] \\
&= 1
\end{align*}

\subsection{Uncertainty of Boss Rooms}
Bosses are incredibly difficult enemies encountered at the end of each Act. Compared to regular monsters and elites, bosses have significantly higher health, deal more devastating attacks, and employ unique mechanics that either severely weaken the player's deck or greatly enhance their own abilities. Each boss has a distinctive portrait displayed at the top of the Spire map, allowing players to strategize their deck accordingly. The location before a boss battle is always a Rest Site, giving the player an opportunity to heal before the fight\footnote{\url{https://slay-the-spire.fandom.com/wiki/Bosses}}.

\begin{itemize}
    \item Act 1 (Floor 16): \texttt{Slime Boss}, or \texttt{The Guardian}, or \texttt{Hexaghost}.
    \item Act 2 (Floor 33): \texttt{Bronze Automaton}, or \texttt{The Champ}, or \texttt{The Collector}.
    \item Act 3 (Floor 50): \texttt{Awakened One}, or \texttt{Time Eater}, or \texttt{Donu and Deca}.
\end{itemize}

\begin{multline*}
H(\text{boss rooms}) = -\sum p_i \log p_i \\
= -\big[3 \times (33.33\% \log 33.33\%)\big] = 1.58
\end{multline*}
\begin{figure}[b]
\centering
\includegraphics[width=0.5\textwidth]{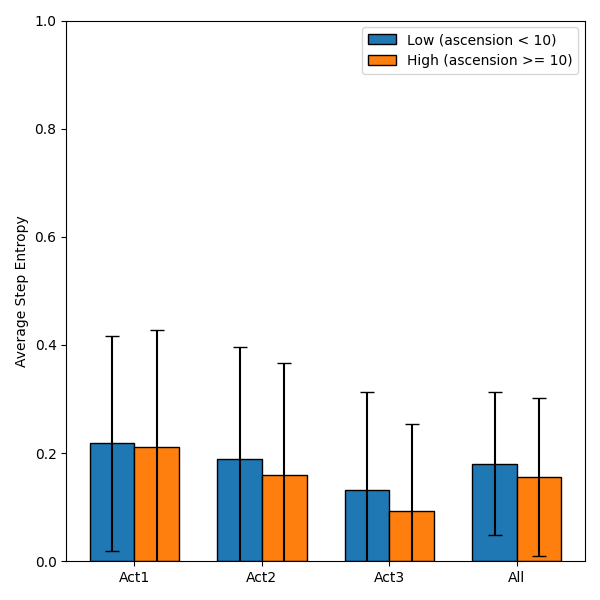}
\caption{Relationship between Average Normalized Entropy in each step and Player Level}
\label{fig:levels}
\end{figure}
\subsection{Uncertainty of Treasure Rooms}
Non-boss chests can be found in a Treasure Room. These chests will give the player one relic from the Common, Uncommon, or Rare relic pool. They will also have a chance to give the player a certain amount of gold\footnote{\url{https://slay-the-spire.fandom.com/wiki/Treasure_Room}}.

\begin{multline*}
H(\text{treasure rooms}) = -\sum p_i \log p_i \\
= -\big[(50\% \log 50\%) + (33\% \log 33\%) + (17\% \log 17\%)\big] = 1.46
\end{multline*}

\subsection{Uncertainty of Shop Rooms}
The Shop is one of the Map Locations, and it is the primary way for the player to spend their Gold. There is no uncertainty in Shop Rooms\footnote{\url{https://slay-the-spire.fandom.com/wiki/Merchant}}.

\noindent
\small
\[
    H(\text{shop rooms}) = 0
\]

\subsection{Uncertainty of Rest Sites}
By default, rest sites only offer a choice between Rest and Smith. More options can be added by specific Relics. Only one option can be chosen at each rest site. There is no uncertainty in Shop Rooms\footnote{\url{https://slay-the-spire.fandom.com/wiki/Rest_Site}}.

\noindent
\small
\[
H(\text{rest sites}) = 0 
\]

\subsection{Uncertainty of Unknown Locations}
When the player visits one of the Unknown Locations, they may encounter an Event, a Monster, a Shop, or a Treasure Room\footnote{\url{https://slay-the-spire.fandom.com/wiki/Unknown_Location}}. Upon entering the first Unknown room in an act, the chances of finding each type of encounter are:

\begin{itemize}
  \item \textbf{Monster}: 10\%, increased by 10\% each time a Unknown room does not contain Monster.
  \item \textbf{Treasure}: 2\%, increased by 2\% each time a Unknown room does not contain Treasure. (With the Deadly Events modifier, the chance is increased by 4\% each time instead.) 
  \item \textbf{Shop}: 3\%, increased by 3\% each time a Unknown room does not contain Shop.
  \item \textbf{Elite}: 20\%, increased by 20\% each time a Unknown room does not contain Elite. (This only applies if using the Deadly Events modifier, starting at floor 6 and above.)
  \item \textbf{Event}: if no other encounter is found.
\end{itemize}

The first unknown location is a random variable \( X \) with 5 possible outcomes:
\[
X = \begin{cases} 
    M & \text{with probability } 10\%, \\ 
    T & \text{with probability } 2\%, \\ 
    S & \text{with probability } 3\%, \\ 
    E & \text{with probability } 20\%, \\ 
    Q & \text{with probability } 65\%.
\end{cases}
\]

The second unknown location is a random variable conditioned on first unknown location:
The conditional probabilities for \( X_2 \) given \( X_1 \) are as follows:

\[
\{X_2 | X_1 = M\} = \begin{cases}
    M & \text{with probability } 10\%, \\
    T & \text{with probability } 4\%, \\
    S & \text{with probability } 6\%, \\
    E & \text{with probability } 40\%, \\
    Q & \text{with probability } 40\%.
\end{cases}
\]

\[
\{X_2 | X_1 = T \} = \begin{cases}
    M & \text{with probability } 20\%, \\
    T & \text{with probability } 2\%, \\
    S & \text{with probability } 6\%, \\
    E & \text{with probability } 40\%, \\
    Q & \text{with probability } 32\%.
\end{cases}
\]

\[
\{X_2 | X_1 = S \} = \begin{cases}
    M & \text{with probability } 20\%, \\
    T & \text{with probability } 4\%, \\
    S & \text{with probability } 3\%, \\
    E & \text{with probability } 40\%, \\
    Q & \text{with probability } 33\%.
\end{cases}
\]

\[
\{X_2 | X_1 = E \} = \begin{cases}
    M & \text{with probability } 20\%, \\
    T & \text{with probability } 4\%, \\
    S & \text{with probability } 6\%, \\
    E & \text{with probability } 20\%, \\
    Q & \text{with probability } 50\%.
\end{cases}
\]

\[
\{X_2 | X_1 = Q \} = \begin{cases}
    M & \text{with probability } 20\%, \\
    T & \text{with probability } 4\%, \\
    S & \text{with probability } 6\%, \\
    E & \text{with probability } 40\%, \\
    Q & \text{with probability } 30\%.
\end{cases}
\]
To calculate the entropy of the sequence \( (X_1, X_2, \ldots, X_{15}) \) using the chain rule, we express the joint entropy as follows:

\begin{multline*}
H(X_1, X_2, \ldots, X_{15}) = \\
H(X_1) + H(X_2 | X_1) + H(X_3 | X_1, X_2) + \cdots + H(X_{15} | X_1, X_2, \ldots, X_{14})
\end{multline*}

The maximum length of paths is $15$ therefore we show the general case in this length, in practice, there's always less than 15 unknown locations in a path.

This formulation allows us to compute the entropy of each variable given the previous ones, leveraging the chain rule for entropy. Given the dependencies we specified, we can simplify the calculations by assuming each \( H(X_i | X_{i-1}) \) depends only on \( X_{i-1} \) (a first-order Markov property). This means we approximate \( H(X_i | X_1, X_2, \ldots, X_{i-1}) \approx H(X_i | X_{i-1}) \).

\subsubsection{Calculation Steps}

\begin{enumerate}
\item  We compute \( H(X_1) \):
    \noindent
    \small
    \[
    H(X_1) = -\sum_{x \in \{M, T, S, E, Q\}} P(X_1 = x) \log P(X_1 = x)
    \]

\item  We compute \( H(X_i | X_{i-1}) \) for \( i = 2, \ldots, 15 \), if appeared in path:
For each outcome \( x \in \{M, T, S, E, Q\} \) of \( X_{i-1} \), we compute:
    
    \begin{multline*}
    H(X_i | X_{i-1} = x) = \\
    -\sum_{y \in \{M, T, S, E, Q\}} P(X_i = y | X_{i-1} = x) \log P(X_i = y | X_{i-1} = x)
    \end{multline*}
Then, we weigh these by \( P(X_{i-1} = x) \):
    \noindent
    \small
    \[
    H(X_i | X_{i-1}) = \sum_{x} P(X_{i-1} = x) H(X_i | X_{i-1} = x)
    \]

\item  We sum up the entropies:
    \noindent
    \small
    \[
    H(X_1, X_2, \ldots, X_{15}) = H(X_1) + \sum_{i=2}^{15} H(X_i | X_{i-1})
    \]
\end{enumerate}






\section{Evaluation}

We used the \texttt{scipy.stats.ttest\_ind} function to perform the t-test, specifically with the argument \texttt{equal\_var=False} to account for unequal variances between the groups. This function returns the test statistic and the p-value, which we used to evaluate the hypothesis. The significance level ($\alpha$) was set at 0.05, which is commonly used in hypothesis testing.

\subsection{Relationship between Normalized Entropy and End Result of the Run}
A significant difference in normalized entropy was observed between the Victory and Defeat groups across all acts. The t-test results revealed significant differences in the averaged normalized entropies for all acts as a whole (\( t = -64.63, p < 0.001 \)). Additionally:
\begin{itemize}
    \item \textbf{Act 1}: \( t = -1.14, p = 0.25 \), not significant, implying that victorious and defeated players do not show different risk-taking behavior in the early game.
    \item \textbf{Act 2}: \( t = -39.70, p < 0.001 \), showing a significant difference.
    \item \textbf{Act 3}: \( t = -87.78, p < 0.001 \), confirming the significant difference.
\end{itemize}

These findings suggest on average runs ending with victory are the runs that embrace the uncertainty in the paths and include paths with higher entropy.

\subsection{Relationship between Normalized Entropy and Player Level}
A significant difference in normalized entropy was observed between lower (levels $\leq$ 10) and higher (levels $>$ 10) skill groups across all acts. The t-test results revealed significant differences in the averaged normalized entropies for all acts as a whole (\( t = 14.12, p < 0.001 \)). Additionally:

\begin{itemize}
    \item \textbf{Act 1}: \( t = 1.93, p = 0.053 \), not significant, implying that skill level does not influence risk-taking behavior in the early game.
    \item \textbf{Act 2}: \( t = 11.36, p < 0.001 \) demonstrating a significant difference.
    \item \textbf{Act 3}: \( t = 13.89, p < 0.001 \) demonstrating a significant difference.
\end{itemize}

These findings suggest on average more skilled players exhibit more calculated and less risky decision-making compared to less experienced players.

\section{Results}

The analysis of normalized entropy across the three acts reveals interesting insights into player behavior and risk-taking tendencies. As shown in Figure \ref{fig:results}, players generally exhibit a broad range of normalized entropy values, reflecting diverse risk preferences and decision-making strategies during gameplay.

The results show that, on average, normalized entropy differs significantly between victorious and defeated runs, with a p-value indicating strong statistical significance. This suggests that risk-taking is associated with favorable outcomes, possibly reflecting the need for bold strategies to succeed in challenging scenarios. Players who take calculated risks may be better positioned to overcome fights.

Further statistical analysis reveals that player skill level, measured by ascension level, influences normalized entropy in Acts 2 and 3, as well as on average across all acts. This indicates ascension level does not significantly affect normalized entropy in Act 1, suggesting that early-game risk-taking behavior is relatively uniform across players regardless of skill.  As shown in Figure \ref{fig:levels}, as players gain more experience or skill, their risk-taking behavior evolves in later stages of the game. Specifically, experienced players may exhibit a more refined balance between risk and caution, adapting their strategies to the increasing complexity and stakes of the game.

\section{Discussion and Conclusion}
This study explored path uncertainty in Slay the Spire --- where procedurally generated maps create diverse paths and choices --- and demonstrated its relationship with player outcomes and skill levels.  The results show that victorious runs are associated with higher normalized entropy, supporting the idea that calculated risk-taking can lead to favorable outcomes. Additionally, skill level significantly influences risk-taking behavior, especially in the later stages of the game as its complexity increases. Future research could expand this analysis to other aspects of randomness in the game or explore its application to similar procedurally generated games.

\begin{acks}
The authors would like to thank Wolfgang Gatterbauer and Javed Aslam their feedback on an earlier version of this work. We would also like to acknowledge the assistance of Wordtune to improve the quality of the writing for this paper.
\end{acks}

\bibliographystyle{ACM-Reference-Format}
\bibliography{refs}

\appendix
\onecolumn
\section{Appendix}
\subsection{Dataset Attributes}
\begin{table*}[h]
\centering
\begin{tabular}{|p{0.05\linewidth}|p{0.25\linewidth}|p{0.15\linewidth}|p{0.45\linewidth}|}
\hline
& \textbf{Attribute} & \textbf{Type} &  \textbf{Description} \\ 
\hline
1 & Victory & \texttt{boolean} & Specifies whether the player achieved victory. \\ 
\hline
2 & Ascension level & \texttt{int} & The ascension level of the player. \\ 
\hline
3 & Seed played & \texttt{string} & The seed value used in the run. \\ 
\hline
4 & Path taken & \texttt{category} & The general path chosen in the run. \\ 
\hline
\end{tabular}
\caption{Variable attributes of interest available in the dataset.}
\label{tab:variable}
\end{table*}

\begin{table*}[h]
\centering
\begin{tabular}{|p{0.05\linewidth}|p{0.25\linewidth}|p{0.15\linewidth}|p{0.35\linewidth}|p{0.05\linewidth}|}
\hline
 & \textbf{Attribute} & \textbf{Type} & \textbf{Description} & \textbf{Value}\\ 
\hline
1 & Is ascension mode & \texttt{boolean} & If the game mode is set to ascension. & True\\ 
\hline
2 & Is trial & \texttt{boolean} & Specifies whether the run is for a trial session. & False\\ 
\hline
3 & Character chosen & \texttt{category} & The character selected for gameplay. & Iron Clad\\ 
\hline
4 & Is daily & \texttt{boolean} & Whether the entry is part of a daily challenge. & False\\ 
\hline
5 & Chose seed & \texttt{boolean} & If a specific seed was chosen for the run. & False\\ 
\hline
6 & Is beta & \texttt{boolean} & If the run corresponds to a beta version. & False\\ 
\hline
7 & Is endless & \texttt{boolean} & If the run is set to endless game mode. & False\\ 
\hline
\end{tabular}
\caption{Fixed Attributes available in the dataset.}
\label{tab:fixed}
\end{table*}

\subsection{Probability Distribution Tables}
This section contains tables detailing the different encounters at each map location. These tables, along with other information on map locations and encounter probabilities, have been sourced from the game's Fandom Wiki\footnote{\url{https://slay-the-spire.fandom.com/wiki/}}.

\begin{table*}[h]
    \centering
    \begin{tabular}{|p{0.15\textwidth}|p{0.60\textwidth}|p{0.15\textwidth}|}
        \hline
        \textbf{Name} & \textbf{Details} & \textbf{Encounter chance} \\
        \hline
        \texttt{Cultist} & & 25\% \\
        \hline
        \texttt{Jaw Worm} & & 25\% \\
        \hline
        \texttt{2 Louses} & Each \texttt{Louse} has 50\% chance of being either red or green. & 25\% \\
        \hline
        \texttt{Small Slimes} & \texttt{Acid Slime} or \texttt{Spike Slime} + \texttt{Acid Slime} or \texttt{Spike Slime} & 25\% \\
        \hline
    \end{tabular}
    \caption{Act 1 First Three Encounters}
    \label{tab:act1_first}
\end{table*}

\begin{table*}[h]
    \centering
    \begin{tabular}{|p{0.20\textwidth}|p{0.50\textwidth}|p{0.20\textwidth}|}
        \hline
        \textbf{Name} & \textbf{Details} & \textbf{Encounter chance} \\
        \hline
        \texttt{Gremlin Gang} & 4 randomly chosen from 2 \texttt{Mad}, 2 \texttt{Sneaky}, 2 \texttt{Fat}, 1 \texttt{Wizard}, and 1 \texttt{Shield Gremlin}. & 6.25\% \\
        \hline
        \texttt{Large Slime} & \texttt{Acid Slime} or \texttt{Spike Slime}. & 12.5\% \\
        \hline
        \texttt{Lots of Slimes} & 3 \texttt{Spike Slime} + 2 \texttt{Acid Slime}. & 6.25\% \\
        \hline
        \texttt{Blue Slaver} & - & 12.5\% \\
        \hline
        \texttt{Red Slaver} & - & 6.25\% \\
        \hline
        \texttt{3 Louses} & Each \texttt{Louse} has 50\% chance of being either red or green. & 12.5\% \\
        \hline
        \texttt{2 Fungi Beasts} & - & 12.5\% \\
        \hline
        \texttt{Exordium Thugs} & Any \texttt{Louse} or \texttt{Acid/Spike Slime} + \texttt{Looter} or \texttt{Cultist} or Any \texttt{Slaver}. & 9.375\% \\
        \hline
        \texttt{Exordium Wildlife} & \texttt{Fungi Beast} or \texttt{Jaw Worm} + Any \texttt{Louses} or \texttt{Acid/Spike Slime}. & 9.375\% \\
        \hline
        \texttt{Looter} & - & 12.5\% \\
        \hline
    \end{tabular}
    \caption{Act 1 Rest of Encounters}
    \label{tab:act1_rest}
\end{table*}

\begin{table*}[h]
    \centering
    \begin{tabular}{|p{0.2\textwidth}|p{0.4\textwidth}|p{0.3\textwidth}|}
        \hline
        \textbf{Name} & \textbf{Details} & \textbf{Encounter chance} \\
        \hline
        \texttt{Spheric Guardian} & - & 20\% \\
        \hline
        \texttt{Chosen} & - & 20\% \\
        \hline
        \texttt{Shelled Parasite} & - & 20\% \\
        \hline
        \texttt{3 Byrds} & - & 20\% \\
        \hline
        \texttt{2 Thieves} & \texttt{Looter} + \texttt{Mugger} & 20\% \\
        \hline
    \end{tabular}
    \caption{Act 2 First Two Encounters}
    \label{tab:act2_first}
\end{table*}

\begin{table*}[h]
    \centering
    \begin{tabular}{|p{0.63\textwidth}|p{0.3\textwidth}|}
        \hline
        \textbf{Name} & \textbf{Encounter chance} \\
        \hline
        \texttt{Chosen and Byrd} & 7\% \\
        \hline
        \texttt{Cultist and Chosen} & 10\% \\
        \hline
        \texttt{Sentry and Spheric Guardian} & 7\% \\
        \hline
        \texttt{Snake Plant} & 21\% \\
        \hline
        \texttt{Snecko} & 14\% \\
        \hline
        \texttt{Centurion and Mystic} & 21\% \\
        \hline
        \texttt{3 Cultists} & 10\% \\
        \hline
        \texttt{Shelled Parasite and Fungi Beast} & 10\% \\
        \hline
    \end{tabular}
    \caption{Act 2 Rest of Encounters}
    \label{tab:act2_rest}
\end{table*}

\begin{table*}[h]
    \centering
    \begin{tabular}{|p{0.15\textwidth}|p{0.55\textwidth}|p{0.2\textwidth}|}
        \hline
        \textbf{Name} & \textbf{Details} & \textbf{Encounter chance} \\
        \hline
        \texttt{3 Darklings} & - & 33.33\% \\
        \hline
        \texttt{Orb Walker} & - & 33.33\% \\
        \hline
        \texttt{3 Shapes} & 3 randomly chosen from 2 \texttt{Repulsors}, 2 \texttt{Spikers}, and 2 \texttt{Exploders}. & 33.33\% \\
        \hline
    \end{tabular}
    \caption{Act 3 First Two Encounters}
    \label{tab:act3_first}
\end{table*}

\begin{table*}[h]
    \centering
    \begin{tabular}{|p{0.24\textwidth}|p{0.46\textwidth}|p{0.2\textwidth}|}
        \hline
        \textbf{Name} & \textbf{Details} & \textbf{Encounter chance} \\
        \hline
        \texttt{4 Shapes} & 4 randomly chosen from 2 \texttt{Repulsors}, 2 \texttt{Spikers}, and 2 \texttt{Exploders}. & 12.5\% \\
        \hline
        \texttt{Maw} & - & 12.5\% \\
        \hline
        \texttt{Spheric Guardian and 2 Shapes} & Shapes are randomly chosen from \texttt{Repulsors}, \texttt{Spikers}, and \texttt{Exploders}. & 12.5\% \\
        \hline
        \texttt{3 Darklings} & - & 12.5\% \\
        \hline
        \texttt{Writhing Mass} & - & 12.5\% \\
        \hline
        \texttt{Jaw Worm Horde} & 3 \texttt{Jaw Worms} & 12.5\% \\
        \hline
        \texttt{Spire Growth} & - & 12.5\% \\
        \hline
        \texttt{Transient} & - & 12.5\% \\
        \hline
    \end{tabular}
    \caption{Act 3 rest of monster encounters}
    \label{tab:act3_rest}
\end{table*}

\begin{table*}[h]
    \centering
    \begin{tabular}{|p{0.28\textwidth}|p{0.2\textwidth}|p{0.2\textwidth}|p{0.2\textwidth}|}
        \hline
        & \textbf{Small Chest} & \textbf{Medium Chest} & \textbf{Large Chest} \\
        \hline
        \textbf{Chance} & 50\% & 33\% & 17\% \\
        \hline
        \multicolumn{4}{|c|}{\textbf{Relic Rewards}} \\
        \hline
        Common & 75\% & 35\% & 0\% \\
        Uncommon & 25\% & 50\% & 75\% \\
        Rare & 0\% & 15\% & 25\% \\
        \hline
        \multicolumn{4}{|c|}{\textbf{Gold Rewards}} \\
        \hline
        Chance & 50\% & 35\% & 50\% \\
        Amount & 23-27 & 45-55 & 68-82 \\
        \hline
    \end{tabular}
    \caption{Chest Rewards and Chances}
    \label{tab:chests}
\end{table*}

\end{document}